\documentclass[twocolumn]{aastex631}
\usepackage{float}
\usepackage{booktabs}
\newcommand\sSFRG{$\text{sSFR}_{100 - 1000 \text{Myr}}$}
\newcommand\sSFRM{$\text{sSFR}_{0 - 100 \text{Myr}}$}
\newcommand\DN{D$_{\text{n}}$4000}

\shorttitle{A simple spectroscopic technique to identify rejuvenating galaxies}
\shortauthors{Zhang et al.}

\graphicspath{{./}{figures/}}

\begin{document}

\title{A simple spectroscopic technique to identify rejuvenating galaxies}

\author[0000-0002-1574-2045]{Junyu Zhang}
\affiliation{Department of Astronomy \& Astrophysics, The Pennsylvania State University, University Park, PA 16802, USA}
\affiliation{Institute for Computational \& Data Sciences, The Pennsylvania State University, University Park, PA 16802, USA}
\affiliation{Institute for Gravitation and the Cosmos, The Pennsylvania State University, University Park, PA 16802, USA}

\author[0000-0002-0682-3310]{Yijia Li}
\affiliation{Department of Astronomy \& Astrophysics, The Pennsylvania State University, University Park, PA 16802, USA}
\affiliation{Institute for Computational \& Data Sciences, The Pennsylvania State University, University Park, PA 16802, USA}
\affiliation{Institute for Gravitation and the Cosmos, The Pennsylvania State University, University Park, PA 16802, USA}

\author[0000-0001-6755-1315]{Joel Leja}
\affiliation{Department of Astronomy \& Astrophysics, The Pennsylvania State University, University Park, PA 16802, USA}
\affiliation{Institute for Computational \& Data Sciences, The Pennsylvania State University, University Park, PA 16802, USA}
\affiliation{Institute for Gravitation and the Cosmos, The Pennsylvania State University, University Park, PA 16802, USA}

\author[0000-0001-7160-3632]{Katherine E. Whitaker}
\affiliation{Department of Astronomy, University of Massachusetts, Amherst, MA 01003, USA}
\affiliation{Cosmic Dawn Center (DAWN), Niels Bohr Institute, University of Copenhagen, Jagtvej 128, København N, DK-2200, Denmark}

\author[0000-0001-6843-409X]{Angelos Nersesian}
\affiliation{Sterrenkundig Observatorium, Universiteit Gent, Krijgslaan 281 S9, 9000 Gent, Belgium}

\author[0000-0001-5063-8254]{Rachel Bezanson}
\affiliation{Department of Physics and Astronomy and PITT PACC, University of Pittsburgh, Pittsburgh, PA 15260, USA}

\author[0000-0002-5027-0135]{Arjen van der Wel}
\affiliation{Sterrenkundig Observatorium, Universiteit Gent, Krijgslaan 281 S9, 9000 Gent, Belgium}

\begin{abstract}
Rejuvenating galaxies are unusual galaxies that fully quench and then subsequently experience a ``rejuvenation" event to become star-forming once more. Rejuvenation rates vary substantially in models of galaxy formation: 10\%-70\% of massive galaxies are expected to experience rejuvenation by $z = 0$. Measuring the rate of rejuvenation is therefore important for calibrating the strength of star formation feedback mechanisms. However, these observations are challenging because rejuvenating systems blend in with normal star-forming galaxies in broadband photometry. In this paper, we use the galaxy spectral energy distribution (SED)-fitting code Prospector to search for observational markers that distinguish normal star-forming galaxies from rejuvenating galaxies. We find that rejuvenating galaxies have smaller Balmer absorption line equivalent widths (EWs) than normal star-forming galaxies. This is analogous to the well-known ``K + A" or post-starburst galaxies, which have strong Balmer absorption due to A-stars dominating the light: in this case, rejuvenating systems have a lack of A-stars, instead resembling ``O - A" systems. We find star-forming galaxies that have H$\beta$, H$\gamma$, and/or H$\delta$ absorption EWs $\lesssim 3$\AA~corresponds to a highly pure selection of rejuvenating systems. Interestingly, while this technique is highly effective at identifying mild rejuvenation, ``strongly" rejuvenating systems remain nearly indistinguishable from star-forming galaxies due to the well-known stellar outshining effect. We conclude that measuring Balmer absorption line EWs in star-forming galaxy populations is an efficient method to identify rejuvenating populations, and discuss several techniques to either remove or resolve the nebular emission which typically lies on top of these absorption lines.
\end{abstract}

\keywords{Galaxy evolution (594); Galaxy formation (595); Green valley galaxies (683); Star formation (1569); Galaxy accretion (575); Post-starburst galaxies (2176)}

\section{Introduction} \label{sec:intro}

It has long been observed that the optical colors of galaxies are bimodal (e.g., \citealt{Strateva_2001,Baldry_2004}). This is true in both the local universe and at higher redshift (e.g., \citealt{Bell_2004,Franzetti_2007, Whitaker_2011, Straatman_2016}). Galaxy populations can be divided by this bimodality into two dominant regions: the ``blue cloud" and ``red sequence". The blue cloud is mainly populated by star-forming galaxies, while the red sequence is mainly populated by quiescent galaxies which show little ongoing star formation. There is also a transition area called the ``green valley" that separates the blue cloud from the red sequence (e.g., \citealt{Strateva_2001,Fang_2012,Salim_2014}). Many studies show the number density of quiescent galaxies increases as the universe gets old (\citealt{Pozzetti_2010, Brammer_2011, Moustakas_2013,Muzzin_2013, Leja_2022}), which suggests that galaxies evolve from being star-forming to quiescent. 

However, some quiescent galaxies may experience ``rejuvenation", with a secondary burst of star formation after a long period of low-level star formation. Through the rejuvenation event, galaxies in the red sequence may be able to return to the green valley and occasionally even to the blue cloud (e.g., \citealt{Rowlands_2017,Chauke_2019}). The nature of this rejuvenation process is still poorly understood, although gas accretion during gas-rich minor mergers and smooth accretion from the intergalactic medium (IGM) are thought to be possible triggers of the rejuvenation process (e.g.,  \citealt{kaviraj2009role, Fang_2012, Dressler_2013, Rowlands_2017}). Understanding the rejuvenation process is crucial because it can help better constrain the wide range of star formation regulation prescriptions in theoretical models of galaxy formation. 

Observations of rejuvenating galaxies suggest that the frequency of rejuvenation is low and possibly redshift-dependent. Rejuvenation appears to happen more frequent in the local universe, with $\sim$10\%-30\% of early-type galaxies at $z < 0.1$ showing evidence of rejuvenation (\citealt{Treu_2005,Schawinski_2007,Donas_2007,Thomas_2010}). Furthermore, there are hints that rejuvenation happens less frequently at high redshift. Recent high-redshift observations suggest that $<$ 1-16\% of high stellar mass quiescent galaxies ($\log M_{\ast} > 11 M_{\sun}$) have experienced past rejuvenation events, depending on the definition of a rejuvenation (\citealt{Belli_2017,Chauke_2019,Tacchella_2022,woodrum_2022}). In addition, \cite{Akhshik_2021} found a rejuvenating galaxy at z = 1.88 and estimated that only about 1\% of massive quiescent galaxies at 1 \textless~z \textless~2 are possibly rejuvenating. However, due to the different methodologies and definitions, it is hard to determine the consistency between these studies, or establish a true global rejuvenation rate.

Rejuvenation rates can be measured more easily in simulations of galaxy formation. \cite{Rodr_guez_Montero_2019} demonstrate in the Simba simulation that the frequency of rejuvenation is much lower than the rate of mergers and quenching events, suggesting it is a sub-dominant process in the quiescent population. The frequency of rejuvenation is found to be higher in both intermediate mass galaxies regardless of redshift and low mass galaxies of low redshift, which is consistent with the aforementioned observational results. \citet{Alarcon_2022} have shown that 40\%-70\% of massive $z = 0$ galaxies experience a rejuvenation event in UniverseMachine \citep{Behroozi_2019}, while the rate in IllustrisTNG \citep{Pillepich_2017,Springel_2017, Nelson_2017} is only $\sim$10\%. We note, however, that since there is no universal quantitative definition of rejuvenation, there is some uncertainty in comparing the fractions of rejuvenation from different studies.

However, finding galaxies that are undergoing rejuvenation, is extremely challenging since rejuvenating galaxies are likely to be highly degenerate with the regular star-forming population in colors from broadband photometry. In addition to the specific star formation rate (sSFR = SFR/$M_{\ast}$), metallicity and dust attenuation can also significantly affect the color of a galaxy, further scrambling the relationship between color and past star formation activity. Moreover, the UV upturn phenomenon caused by hot evolved helium-burning stars can make galaxies with low recent star formation lie in the blue side in FUV-V color \citep{Jeong_2009}, which further complicates the interpretation of galaxy color at low redshift. We caution that the techniques used to estimate rejuvenation frequency vary dramatically in each of the aforementioned observational studies, and an accurate observational census across redshift will require a consistent approach. This is the goal of this paper.

In the past decade, numerous techniques have been created to separate star-forming and quiescent galaxy populations. Rest-frame color-color diagrams, notably the UVJ diagram \citep{Williams_2009} and the NUVrK diagram \citep{Arnouts_2013}, are widely employed to distinguish between quiescent and star-forming galaxies, since they are able to place approximate constraints on sSFR (e.g., \citealt{Wang_2017,Leja_2019b}). Additionally, spectroscopic features such as \DN{} and H$\delta$ absorption line equivalent width (EW) are utilized to diagnose the age of stellar populations and to help determine if galaxies have experienced recent star bursts (\citealt{Kauffmann_2003,Le_Borgne_2006,kauffmann2014quantitative,Maltby_2016,Wu_2018}). These spectral diagnostics can be used to identify types of recent star formation such as post-starburst galaxies \citep{Suess_2022}, and so in this work we investigate whether such techniques can be also used to differentiate the rejuvenating population from the normal star-forming population. In this paper, we employ the Bayesian inference machine \texttt{Prospector} (\citealt{Leja_2017, ben_johnson_2022_6192136, 2021ApJS..254...22J}) to simulate SEDs of rejuvenating and star-forming populations at $z = 0.7$ and search for the possible observational characteristics of the rejuvenating population.

The paper is structured as follows. In Section \ref{sec: model} and \ref{sec: EW}, we describe the SED model for generating simulated galaxy populations and how we measure the EW and \DN. In Section \ref{sec: create}, we generate the model rejuvenating and star-forming populations. In Section \ref{sec:result}, we present the observational characteristics of the rejuvenating population we find in our simulated populations, and propose a new spectroscopic identification criteria for rejuvenating galaxies, while also identifying the contaminants. In Section \ref{sec:disc}, we discuss the advantages and limitations of our technique. Finally, in Section \ref{sec:highlight}, we summarize our conclusions.

\section{SED Modeling} \label{sec: model}
For this work, we employ the \texttt{Prospector} Bayesian inference framework (\citealt{Leja_2017, ben_johnson_2022_6192136, 2021ApJS..254...22J}) and the Flexible Stellar Population Synthesis (FSPS; \citealt{Conroy_2009, Conroy_2010}) code to simulate galaxy SEDs. We adopt the physical model Prospector$-\alpha$ described in \citet{Leja_2017,Leja_2019a}, which uses isochrones based on MIST stellar tracks \citep{Choi_2016, Dotter_2016}. The physical model defines fixed and free parameters for generating the SED, plus their associated priors. Model SEDs of star-forming and rejuvenating populations are generated by drawing from the priors for free parameters in this model. Here we briefly introduce the features included in this model with a special emphasis on the changes made for generating realistic populations of normal star-forming and rejuvenating galaxies. The detailed model parameters and priors are described in \cite{Leja_2019a}.

The Prospector$-\alpha$ model includes a two-component dust attenuation model with a flexible attenuation curve, dust emission powered via energy balance, and free stellar metallicity. This model also includes a 7-bin nonparametric star formation history (SFH). The nonparametric SFH is necessary for this work as it allows a galaxy to increase in SFR after an initial period of quiescence, making it possible to create rejuvenating galaxies in the simulation.

We make a few changes to the Prospector$-\alpha$ model to remove the effects of spectral renormalization and emission lines, and simplify the definition of rejuvenating population. First, in our modified Prospector$-\alpha$ model, we fix the model redshift to 0.7 to make it consistent with the redshift of the LEGA-C data \citep{van_der_Wel_2016}, since we aim to use the LEGA-C data to verify our results in the future. Furthermore, since the total mass formed mostly affects the normalization of the SEDs and we do not want to select rejuvenating galaxies from their masses, we fix the total mass formed at $10^{10}$ $M_{\odot}$ to remove the impact of a variable mass on the model SEDs. In addition, we combine the first two fixed age bins (0 - 30 Myr and 30 - 100 Myr) to simplify our calculation of the average sSFR in the last 100 Myr. This means we do not marginalize over any rapid changes in star formation rate in the last 100 Myr.

We also turn off nebular emission, because nebular emission in galaxies is often difficult to interpret due to complex dust distributions and uncertain ionization sources. This means we are searching for observational signatures of rejuvenation which are solely stellar in nature. We discuss the effects of this change further in Section \ref{sec:disc}. Gas metallicity is also no longer a free parameter since nebular emission is off. Other parameters in Prospector$-\alpha$ model such as dust attenuation are free and follow the prior described in \cite{Leja_2019a}. Table \ref{table1} lists all free parameters and their associated priors for the modified Prospector$-\alpha$ physical model used in this paper.

\begin{table*}[htp]
\huge
    \centering
    \caption{Free Parameters and Their Associated Priors for the modified Prospector$-\alpha$ Physical Model}
    \label{table1}
    \renewcommand\arraystretch{1.5}
    \resizebox{\textwidth}{20mm}{
    \begin{tabular}{lll}
    \hline \hline
    Parameter     & Description & Prior  \\
    \hline
    log($Z/Z_{\odot}$)     &  Stellar metallicity & TopHat: min = -1.0, max = 0.19 \\
    log(SFR ratios) & Ratio of the SFRs in adjacent bins of the six-bin nonparametric SFH (five parameters total) & Student's $t$-distribution with mean = 0, $\sigma$ = 0.3 and $\nu$ = 2 \\
    $\hat{\tau}_{2}$ & Diffuse dust optical depth &  Clipped normal: min = 0, max = 2, mean = 0.3, $\sigma$ = 1 \\
    $\hat{\tau}_{1}$ & Birth-cloud dust optical depth &  Clipped normal in ($\hat{\tau}_{1}$ / $\hat{\tau}_{2}$): min = 0, max = 2, mean = 1,  $\sigma$ = 0.3 \\
    n & Power-law modifier to the shape of the \cite{Calzetti_2000} attenuation curve & TopHat: min = -1.0, max = 0.4\\
    $f_{AGN}$ & AGN luminosity as a fraction of the galaxy bolometric luminosity & Log-uniform: min = $10^{-5}$, max = 3\\
    $\tau_{AGN}$ & Optical depth of AGN torus dust & Log-uniform: min = 5, max = 150 \\
    \hline
    \end{tabular}}
\end{table*}

\section{Measurements of spectral features} \label{sec: EW}
As in the identification of post-starburst (or ``K + A") galaxies, Balmer absorption lines EW and \DN~are promising diagnostics for the identification of variable SFRs. Hence, we prepare to investigate their potential in distinguishing rejuvenating galaxies using our simulated populations. We describe here how we measure Balmer absorption line EWs and \DN.

We measure the H$\beta$, H$\gamma$ and H$\delta$ EWs in our simulated populations using the following steps. First, we fit the spectra from 4000 \,\AA{} to 5000 \,\AA{} with a $4^{th}$ degree polynomial to determine the continuum (i.e., $C(\lambda)$) near H$\beta$, H$\gamma$ and H$\delta$ lines. We mask 4071-4131\,\AA, 4290-4370\,\AA, 4600-4750\,\AA{} and 4831-4891\,\AA, since the spectra usually show strong absorption lines at these wavelengths. As the next step, we subtract the continuum from the spectra and fit the difference with a Voigt profile (i.e., $\mathrm{Voigt}(\lambda)$), which is the convolution of a Lorentz profile and a Gaussian profile, and is frequently applied to examine the spectral broadening of hydrogen Balmer lines (e.g., \citealt{luque2003experimental,tremblay2009spectroscopic,falcon2015laboratory, Wu_2022}). Finally, we integrate $\frac{\mathrm{Voigt}(\lambda)}{C(\lambda)}$ to get the EW,

\begin{equation}
    EW = -\int \frac{\mathrm{Voigt}(\lambda)}{C(\lambda)} \, d\lambda.
\end{equation}
By doing the profile fitting, we can make sure our EW measurements consistent.

For the measurement of \DN, we adopt the \cite{Balogh_1999} definition: the ratio between the average flux density $F_\nu$ between 4000 and 4100 \AA{} and that between 3850 and 3950 \AA. 

\section{Creating a Simulated Star-forming \& Rejuvenating Population} \label{sec: create}

To better understand how star-forming and rejuvenating galaxies behave differently in photometry and spectra, we randomly draw model SEDs from the priors of 11 free parameters from our modified Prospector$-\alpha$ model discussed in Section \ref{sec: model}, and then define rejuvenating and normal star-forming galaxies by their sSFRs in different age bins. We simulate 25,000 rejuvenating galaxies and 25,000 star-forming galaxies in total based on the following criteria:

\begin{itemize}
    \item Normal star-forming galaxy: \\
    $\text{sSFR}_{100 - 1000 \text{Myr}} > 10^{-11} \text{yr}^{-1}$; \\ $\text{sSFR}_{0 - 100 \text{Myr}} > 10^{-10} \text{yr}^{-1}$
    \item Rejuvenating galaxy:\\
    $\text{sSFR}_{100 - 1000 \text{Myr}} < 10^{-11} \text{yr}^{-1}$; \\ $\text{sSFR}_{0 - 100 \text{Myr}} > 10^{-10} \text{yr}^{-1}$.
\end{itemize}
where \sSFRM~and \sSFRG~stand for the sSFR averaged over the most recent 100 Myr and the sSFR averaged over 100 Myr to 1000 Myr respectively. In other words, all objects we have drawn are currently star-forming, but rejuvenating ones were quiescent 100-1000 Myr ago.

\begin{figure} 
    \centering
    \includegraphics[width=0.45\textwidth]{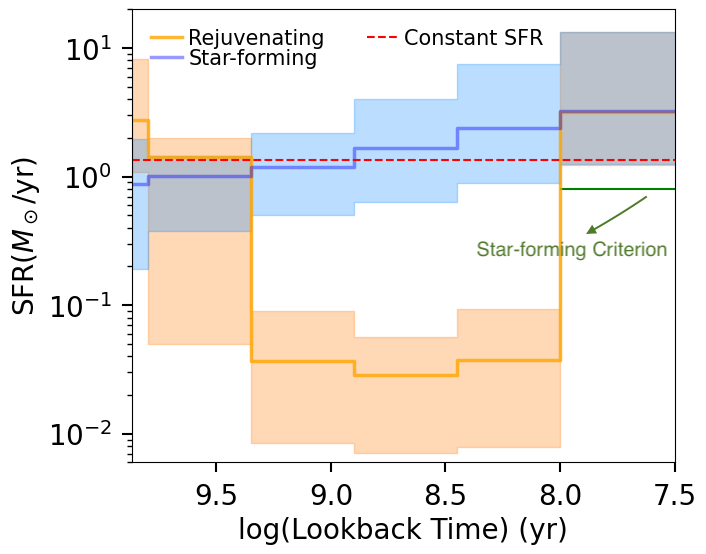}
    \caption{The median SFHs of the simulated star-forming population (blue line) and rejuvenating population (orange line). The light blue and light orange region represent the 16\% to 84\% range of the SFR in each age bin. The rejuvenating population is reweighted by the distribution of \sSFRM~of star-forming population and therefore, the median SFRs of two populations are matched with each other in the 0-100 Myr age bin. The red dashed line represents a constant SFH and the green line represents our star-forming criterion in the most recent age bin. Since we require star-forming galaxies to have a higher SFR than the green line, the resulting median star-forming SFH rises with cosmic time.}
    \label{figure: 1}
    \end{figure}

In Figure \ref{figure: 1} we show the median SFHs of simulated star-forming and rejuvenating populations. The light blue and light orange region represent the 16\% to 84\% range of the SFR of the star-forming and rejuvenating populations in each age bin, respectively. Unlike the normal star-forming population, the rejuvenating population has an ``U-shape" SFH which indicates that they return from being quiescent to star-forming. 

It is important to note that in our simulation the rejuvenating population is undergoing a stronger burst of star formation on average (i.e., higher \sSFRM) compared to the normal star-forming population. This difference is expected and can be explained by our parameter settings: as we fix the total stellar mass formed in the model and set the upper limit on \sSFRG~of rejuvenating galaxies, the rejuvenating population is forced to have high \sSFRM~to meet the total stellar mass formed. However, we do not want to select rejuvenating galaxies based on their higher sSFRs, as that would render this exercise meaningless. To make a fair comparison, we instead assume that the rejuvenating population follows the similar \sSFRM~distribution as the normal star-forming population. For this purpose, we reweigh the rejuvenating samples by drawing rejuvenating samples based on weights from the distribution of normal star-forming samples. Hence, the distribution of rejuvenating samples' \sSFRM~is consistent with that of the normal star-forming sample in the simulation. This reweighing can be seen in the rightmost age bin of Figure \ref{figure: 1}, where the median SFRs of rejuvenating and star-forming population coincide with each other. Note that in the later sections of this paper, the simulated rejuvenating population always refers to the sSFR-weighted simulated rejuvenating population, unless otherwise specified. 

It is noteworthy that the model star-forming population has a median rising SFH in Figure \ref{figure: 1}. To illustrate the reason, we draw a constant SFH (red) and the star-forming criterion (green) in the most recent age bin. Since we require star-forming galaxies to have a higher SFR than the green line, the resultant median SFH which meets this criterion rises with time.

\section{Results} \label{sec:result}
\subsection{The Observational Characteristics of Rejuvenating Galaxies}\label{subsec: spectral}
In this section, we use the model star-forming and rejuvenating populations to search for observational characteristics to separate the two. First, we look for differences in the broadband colors. 

\begin{figure}[htb]
    \centering
    \includegraphics[width=0.45\textwidth]{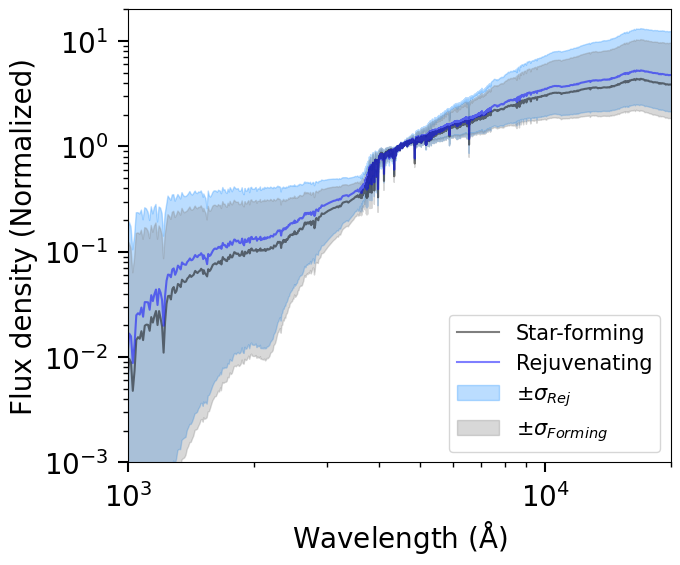}
    \caption{The median spectra of star-forming (black) and rejuvenating (blue) populations, normalized by the average flux density between 4000\,\AA{} and 4400\,\AA{} of each spectra. The grey and light blue regions represent the 16\% to 84\% range of the normalized flux density of star-forming and rejuvenating populations respectively. This illustrates the extreme challenge of selecting rejuvenating systems by their photometry colors alone}
    \label{figure: 2}
    \end{figure}

In Figure \ref{figure: 2} we show the median spectra of the simulated star-forming and rejuvenating populations. For a  better comparison, the median spectra were normalized by the average flux density between 4000\,\AA{} and 4400\,\AA. The grey and light blue regions represent the 16\% to 84\% range of the normalized flux density of star-forming and rejuvenating populations respectively. It can be seen that the simulated rejuvenating and star-forming populations have almost identical range of broadband colors within $\pm 1\sigma$ region, which implies the color difference is not sufficient to differentiate rejuvenating galaxies from normal star-forming galaxies. 

This result makes sense and can be explained in two ways. First, the broadband colors of galaxies are determined not only by their sSFR, but also their dust, metallicity and longer-term variations in their SFHs (see, e.g., \citealt{Conroy_2013} and references therein). Thus, the effect of sSFR on the broadband colors of rejuvenating galaxies can be washed out by the variations of other parameters. Second, it is hard to tell from broadband colors alone whether a galaxy in the ``green valley" is undergoing quenching or rejuvenation. As we mentioned before, rejuvenating galaxies are likely to be in green valley (see, e.g., \citealt{Chauke_2019,Akhshik_2021}). However, when the star-forming galaxies have their star formation quenched, they will also pass through the green valley on their way to becoming completely quiescent. Hence, those recently quenching galaxies can have indistinguishable broadband colors from rejuvenating galaxies. 

\begin{figure*}
  \centering
  \begin{tabular}{@{}c@{}}
    \includegraphics[width=0.67\linewidth]{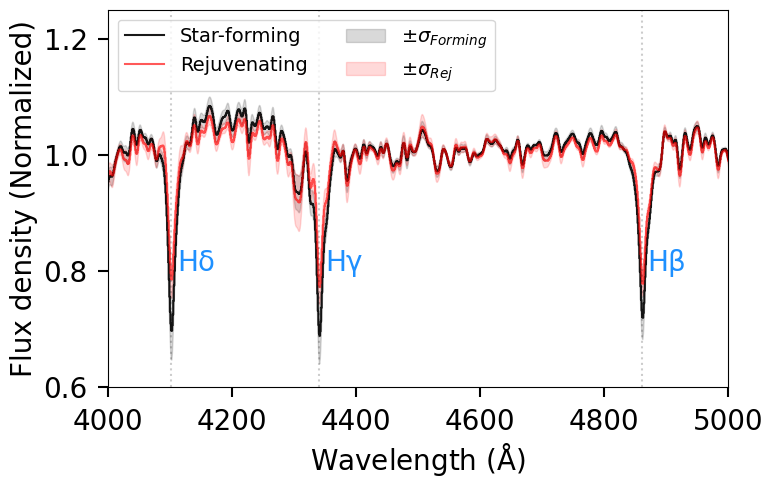}  
  \end{tabular}

  \vspace{\floatsep}

  \begin{tabular}{@{}c@{}}
    \includegraphics[width=\linewidth]{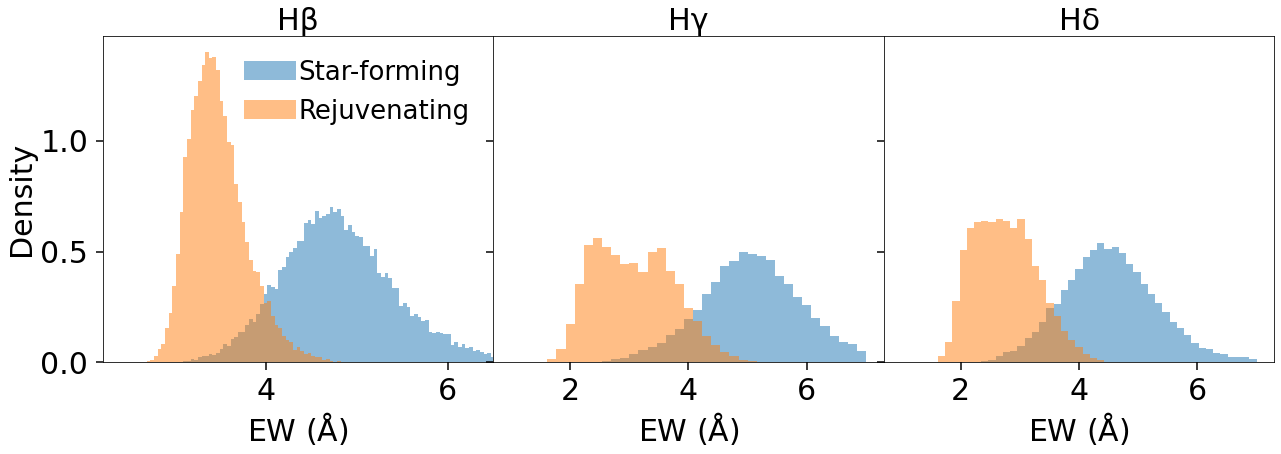}
  \end{tabular}

  \caption{(Top) The median spectra of rejuvenating (red) and star-forming (black) populations from 4000\,\AA{} to 5000\,\AA{} with H$\beta$ $\lambda$4861\,\AA, H$\gamma$ $\lambda$4340\,\AA{} and H$\delta$ $\lambda$4101\,\AA{} marked. To normalize the spectra, we divide each spectrum by their continuum and then take the median value. The light red and grey regions represent the 16\% to 84\% range of the normalized flux density of rejuvenating and star-forming populations respectively. We use a Gaussian filter with $\sigma = 3$ pixels to smooth the spectra for presentation purposes}. (Bottom) Histograms of EW of H$\beta$, H$\gamma$ and H$\delta$ for rejuvenating (orange) and star-forming (blue) populations. The rejuvenating sample is sSFR-weighted as described in Section \ref{sec: create}.
  \label{figure: 3}
\end{figure*}

Figure \ref{figure: 2} demonstrates that broadband colors alone are not able to identify the rejuvenating population, so we now turn to spectra. In the top panel of Figure \ref{figure: 3}, we show the median spectra of rejuvenating and star-forming populations from 4000\,\AA{} to 5000\,\AA. To highlight the differences in absorption lines between the two populations, we normalize spectra by dividing their individual continua before taking the median. It can be seen that the rejuvenating population shows relatively weaker Balmer absorption lines compared to the star-forming population. The physical driver of the Balmer absorption line EWs difference will be further discussed in Section \ref{subsec: SSP}. Here we draw an analogy with post-starburst galaxies. These ``E + A" or ``K + A" galaxies have a significant drop in SFR $\sim$ 100 - 500 Myr ago, and show strong Balmer absorption lines which indicates strong recent star formation but weak ongoing star formation (\citealt{Zabludoff_1996,Dressler_1999,Goto_2005,Brown_2009,French_2015,Suess_2022}). Since the rejuvenating population is currently star-forming but was nearly dead between 100 Myr and 1 Gyr, it is expected to contain fewer A-type stars (which have the strongest Balmer absorption lines) relative to the star-forming population. Therefore, it is reasonable that the spectra of rejuvenating galaxies are likely to demonstrate features of ``O - A" galaxies such as weak Balmer absorption lines.

However, the key question is whether this difference is strong enough to differentiate individual rejuvenating galaxies from normal star-forming galaxies. The bottom row of Figure \ref{figure: 3} shows the EW histograms of H$\beta$, H$\gamma$ and H$\delta$ for the rejuvenating and star-forming populations. In agreement with the top panel, the rejuvenating population shows small EWs of Balmer absorption lines, while the star-forming population exhibits a broad distribution centered at large EWs. Figure \ref{figure: 3} suggests that we can select rejuvenating galaxies based on their small Balmer absorption line EWs, even though the modest overlapping region may contaminate the selection. Therefore, we conclude here that weak Balmer absorption lines is one observational characteristic of the rejuvenating population that can differentiate them from the normal star-forming population in our simulated galaxies. Contaminants and Balmer EW-based criteria to distinguish star-forming and rejuvenating populations will be further discussed in Section \ref{subsec: Contaminants}.

It is worth noting that other Balmer lines also have distinguishing power and we choose these three Balmer lines for observational reasons only: they are all strong and within the LEGA-C spectral window. By using three Balmer lines we can also beat down observational errors of an individual EW measurement. This point is further discussed in Section \ref{subsec: Contaminants}.

\subsection{Why Balmer Absorption Helps In Identifying Rejuvenating Galaxies} \label{subsec: SSP}
\begin{figure*}
    \centering
    \includegraphics[width=0.9\textwidth]{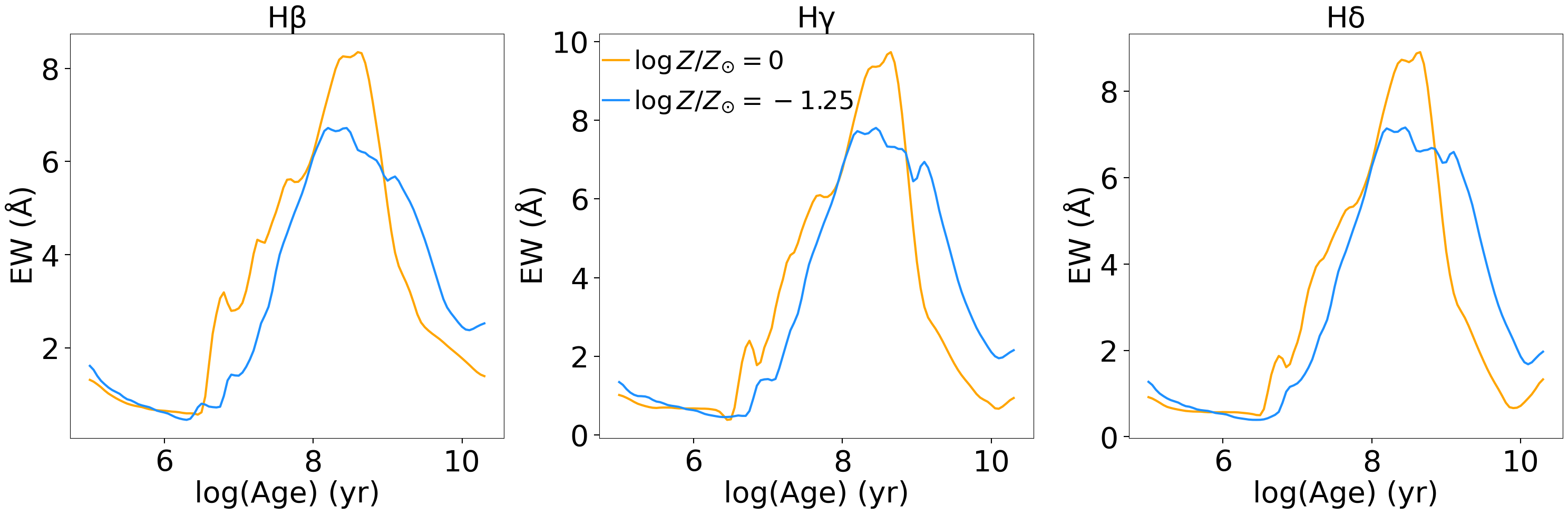}
    \caption{Time evolution of EW of H$\beta$, H$\gamma$ and H$\delta$ (from left to right) at a given metallicity. The blue and orange lines represent $\log Z/Z_\sun$ = -1.25 and $\log Z/Z_\sun$ = 0, respectively.} 
    \label{figure: 4}
    \end{figure*}

In this section, we employ simple stellar populations (SSPs), which are coeval stellar population with a homogeneous metallicity and abundance pattern (e.g., \citealt{Conroy_2013}), to explore the dependence of EW of Balmer absorption lines on age and metallicity. In Figure \ref{figure: 4} we show the evolution of the EWs of H$\beta$, H$\gamma$ and H$\delta$ over time with two sets of stellar metallicities. We only choose two metallicities here, one solar and the other sub-solar, to span a reasonable range of metallicities; super-solar looks almost identical to solar \citep{Mannucci_2010}.

Figure \ref{figure: 4} reveals two important pieces of information. First, we note that the EWs of all three Balmer absorption lines peak in the age range of approximately 100 Myr to 1 Gyr. This result nicely explains why, in Figure \ref{figure: 3}, our model star-forming galaxies have larger Balmer absorption line EWs compared to model rejuvenating galaxies. Due to the continuing strong star formation, star-forming galaxies have more stars with intermediate age (i.e., 100-500 Myr) in Figure \ref{figure: 4}. Hence, they show stronger Balmer absorption lines in spectra. In contrast, rejuvenating galaxies do not have that population of stars and so have weaker Balmer absorption lines.

Another key result we obtain from Figure \ref{figure: 4} is that metallicity generally does not affect our selection via Balmer absorption line EW. For these two stellar populations with different metallicities, their Balmer absorption lines EWs both approximately peak in the same age range, although the lower metallicity results in a weaker peak. Hence, while it is effective to distinguish rejuvenating galaxies regardless of their metallicities through Balmer absorption lines EW, lower metallicity may lead to more contaminants.

\subsection{Efficiency of Detecting Rejuvenating Galaxies} \label{subsec: Contaminants}

\begin{figure*}[htb]
    \centering
    \includegraphics[width=0.9\textwidth]{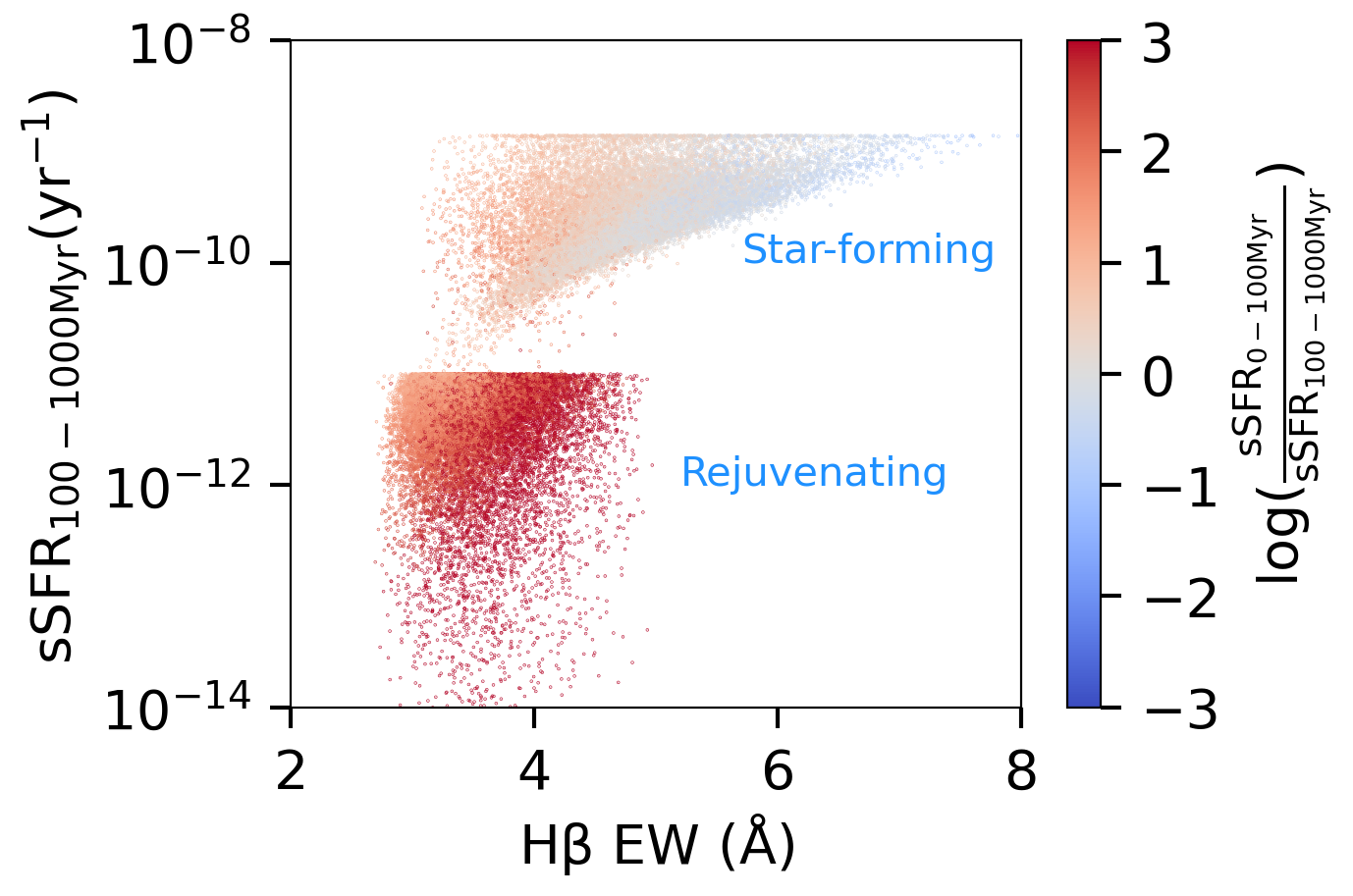}
    \caption{EW of H$\beta$ versus \sSFRG~for rejuvenating and star-forming populations. Points are color-coded by their ratio of \sSFRM~to \sSFRG. We can identify galaxies experiencing mild rejuvenation by applying a reasonable cut in the EW, but galaxies experiencing strong rejuvenation start to blend in with the star-forming galaxy population}
    \label{figure: 5}
    \end{figure*}

Although Figure \ref{figure: 3} shows that relatively weak Balmer absorption lines are a signature of the rejuvenating population, it is important to understand the source of the contaminants in the bottom panels of Figure \ref{figure: 3}. In Figure \ref{figure: 5}, we show EW of H$\beta$ versus \sSFRG~for the rejuvenating and star-forming populations. Objects are color-coded by their ratio of \sSFRM~to \sSFRG, which is referred to as the strength of rejuvenation. As we indicate in Figure \ref{figure: 5}, due to the sSFR cuts we use to classify rejuvenating and star-forming populations, these two different populations are naturally separated. 

\begin{figure*}[htb]
    \centering
    \includegraphics[width=0.9\textwidth]{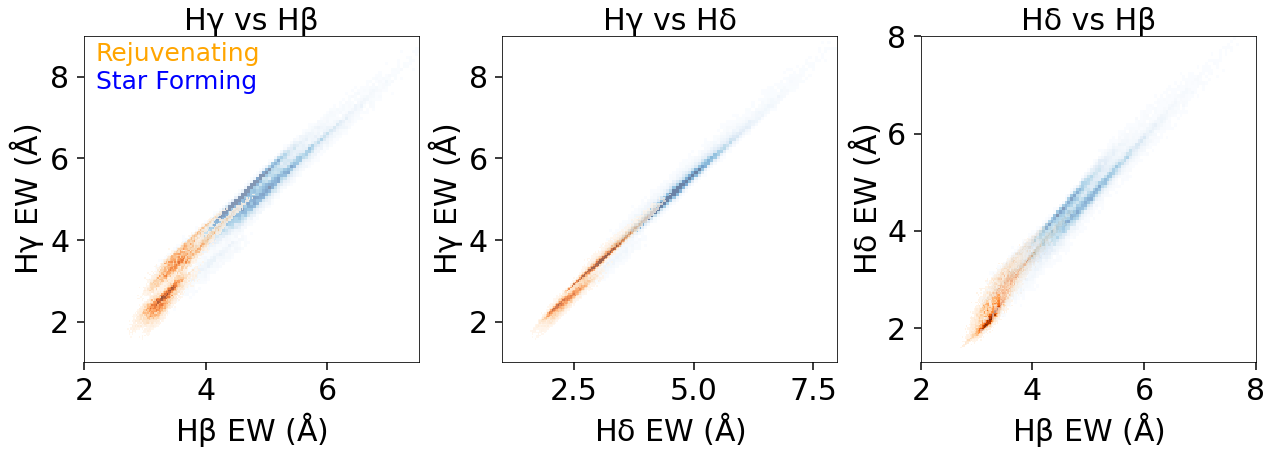}
    \caption{(From left to right) The density plot of H$\gamma$ vs H$\beta$, H$\gamma$ vs H$\delta$ and H$\delta$ vs H$\beta$ EWs for rejuvenating (orange) and star-forming populations (blue). This figure shows that different Balmer absorption lines have the same information. This is due to the fact that they are all Balmer absorption lines, and so are governed by analogous physics and respond similarly to changing stellar age. In this way, it is possible to mitigate the observational error by measuring multiple Balmer absorption lines and averaging them.}
    \label{figure: 6}
    \end{figure*}
    
There are two important notes about our construction of Figure \ref{figure: 5}: firstly, we do not use the sSFR-weighted rejuvenating population in this plot, since the distribution of \sSFRM~for the rejuvenating population is implicitly shown by combining the y-axis and the color-coding of the plot. Secondly, we examine the other two Balmer absorption lines and find that they give the same information. Therefore, we only draw the plot for H$\beta$ here, but the results for H$\gamma$ and H$\delta$ are similar. The reason why all three Balmer absorption lines contain the same information is explained later in this section.

It is interesting that when considering the strength of rejuvenation (indicated by the sSFR ratio), the EWs of the rejuvenating and star-forming populations show opposite trends. As the strength of rejuvenation increases, the EW of H$\beta$ for the rejuvenating population goes up while that of the star-forming population decreases. This is mainly due to the criteria we put on \sSFRG. When the sSFR ratio increases for the star-forming population, that means they are closer to quiescent at 100-1000 Myr, thus they start to look like rejuvenating galaxies. However, when it comes to the rejuvenating population, a larger sSFR ratio means that they have a lot more recent SFR of young stars which outshine the older stars. Increasing the ratio adds more sources of EWs. Therefore, for the rejuvenating population, stronger rejuvenation means larger EWs for Balmer absorption lines.

Figure \ref{figure: 5} can be thought of as the key figure of this paper since it demonstrates both detectable rejuvenating galaxies and the most common contaminant population. We here adopt similar definitions as \cite{Alarcon_2022} for strong and mild rejuvenating galaxies. We use $\log(\frac{{\text{\sSFRM}}}{\text{\sSFRG}}) > 2$ to define strong rejuvenating galaxies and $\log(\frac{\text{\sSFRM}}{\text{\sSFRG}}) > 1~\& < 2$ to define mild rejuvenating galaxies. The EW distribution of the rejuvenating population shows that strong rejuvenating galaxies have relatively large EWs and resemble star-forming galaxies with slightly weaker Balmer absorption lines, which are likely themselves experiencing mild rejuvenation. This suggests that the Balmer EW selection is most efficient at identifying mild rejuvenating galaxies. Additionally, the distribution of star-forming population indicates the contaminants which overlap with the rejuvenating population in the bottom panel of Figure \ref{figure: 3}, are mostly normal star-forming galaxies undergoing weak rejuvenation and were not quiescent between 100 Myr and 1000 Myr. Finally, the rejuvenating population has a smaller lower bound in EW compared to the star-forming population. This means that despite the presence of contaminants, we can still find a parameter space which is solely occupied by the rejuvenating population in our simplified simulation, under the assumption that the star-forming population and the quiescent population can be easily differentiated by other techniques such as rest-frame color-color diagrams. This is consistent with the conclusion we draw from Figure \ref{figure: 3}.

In Figure \ref{figure: 6}, we show the interrelationship of the EW of H$\beta$, H$\gamma$ and H$\delta$ for rejuvenating and star-forming populations. The color coding represents the density of each population. Figure \ref{figure: 6} demonstrates that different Balmer absorption lines show similar contrast between the rejuvenating and star-forming galaxies. Since they are all Balmer absorption lines, they are governed by analogous physics and, therefore, react similarly to the changing stellar age. For this reason, it is feasible to lower the observational uncertainty by combining multiple Balmer absorption lines.

\subsection{H\texorpdfstring{$\delta$}{} vs \texorpdfstring{\DN}{}} \label{subsec:Dn4000}
A popular way to categorize galaxies is via the \DN-H$\delta$ plane (e.g., \citealt{Kauffmann_2003,Le_Borgne_2006, kauffmann2014quantitative, Maltby_2016,Wu_2018, Dhiwar}). To investigate how different types of galaxy populations are located in the \DN-H$\delta$ plane, we also include model quiescent and post-starburst populations here for comparison purposes. We do not include the model quiescent and post-starburst populations in previous sections, since we already have effective methods such as rest-frame color-color diagrams to separate recent star-forming from quiescent galaxies and post-starburst galaxies. The quiescent galaxy population and post-starburst galaxy population are defined as:
\begin{itemize}
    \item Quiescent galaxy: \\
    $\text{sSFR}_{0 - 100 \text{Myr}} < 10^{-11} \text{yr}^{-1}$;\\
    $\text{sSFR}_{100 - 1000 \text{Myr}} < 10^{-10} \text{yr}^{-1}$.
    \item Post-starburst galaxy:\\
    $\text{sSFR}_{0 - 100 \text{Myr}} < 10^{-11} \text{yr}^{-1}$;\\
    $\text{sSFR}_{100 - 1000 \text{Myr}} > 10^{-10} \text{yr}^{-1}$; \\  
\end{itemize}
This means that we allow quiescent galaxies transit from the green valley between 100 Myr and 1 Gyr in lookback time, i.e. they can quench recently.

\begin{figure}
    \centering
    \includegraphics[width=0.45\textwidth]{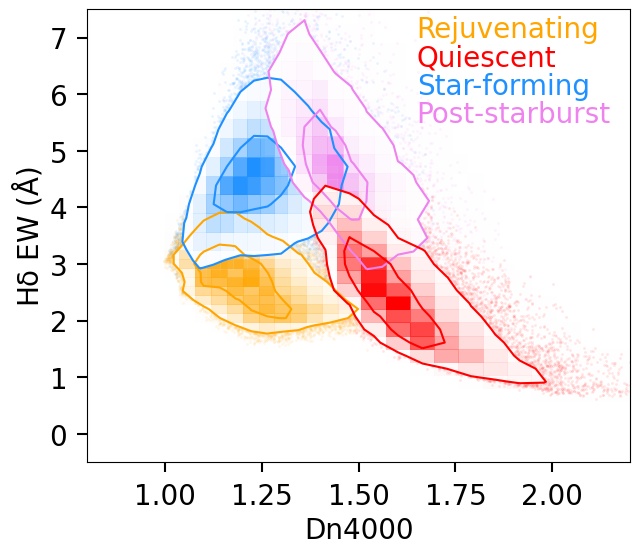}
    \caption{The distribution of model star-forming (blue), rejuvenating (orange), quiescent (red), and post-starburst (violet) populations on the \DN-H$\delta$ plane. The two contours in each population represent $1\sigma$ and $2\sigma$ respectively.}
    \label{figure: 7}
\end{figure}

Figure \ref{figure: 7} shows model star-forming, rejuvenating, quiescent galaxies, and post-starburst galaxies on the \DN-H$\delta$ plane and the two contours in each population represent $1\sigma$ and $2\sigma$ regions respectively. We weight our model rejuvenating population by the \sSFRM~in this contour plot. Figure \ref{figure: 7} demonstrates that the rejuvenating population predominantly lives in the bottom-left corner of the \DN-H$\delta$ plane in our simulation.

Figure \ref{figure: 7} depicts the top-left corner of the \DN-H$\delta$ plane as being populated by model star-forming galaxies, whereas the bottom-right corner is populated by model quiescent galaxies. This is consistent with Figure 6(b) of \citet{Wu_2018} result based on the LEGA-C galaxies. Furthermore, Figure \ref{figure: 7} shows the simulated rejuvenating galaxies have a small \DN. This is due to their strong star formation over the past 100 Myr. On the contrary, in Figure \ref{figure: 7}, post-starburst galaxies exhibit strong Balmer absorption lines because of their high SFRs $\sim$ 500 Myr ago, but due to their weak current star formation, they also show relatively larger \DN. As a result, Figure \ref{figure: 7} suggests that it is possible to employ reasonable cuts on the \DN-H$\delta$ plane to separate rejuvenating galaxies from other galaxy populations.

\subsection{Verification with the LEGA-C survey}
Here we verify that such a population of rejuvenating galaxies as suggested by our model may exist in the data from the LEGA-C survey \citep{van_der_Wel_2016,van_2021}. The LEGA-C survey is a spectroscopic survey observed $\sim$ 4,000 galaxies in the COSMOS field at a redshift range of 0.6 $<$ z $<$ 1.0 with VLT/VIMOS. Each galaxy in the survey was observed for $\sim$ 20 hours, resulting in spectra with $S/N \sim 20 $\AA$^{-1}$ and resolution $R \sim$ 3500. Such a high resolution allows us to resolve emission within absorption lines. 

\begin{figure}
    \centering
    \includegraphics[width=0.45\textwidth]{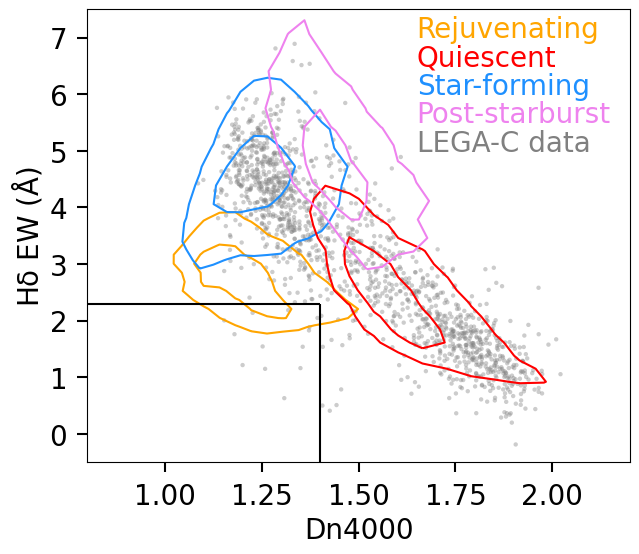}
    \caption{The distribution of 1,429 LEGA-C galaxies on the \DN-H$\delta$ plane, after removing AGNs and low $S/N$ (i.e. average $S/N < 10$ \AA$^{-1}$) sources. The black window represents the proposed selection criteria: H$\delta$ EW $<$ 2.3 \AA~\& \DN~ $<$ 1.4. We also overplot the contours of Figure \ref{figure: 7} for the comparison purpose.}
    \label{figure: 8}
\end{figure}

Since many sources in the LEGA-C survey do not have rest-frame coverage of H$\beta$ or H$\gamma$, it is difficult to improve the purity of our selection by combining the EW criteria of multiple Balmer absorption lines. Therefore, we instead examine how LEGA-C galaxies are distributed on the \DN-H$\delta$ plane. Based on Figure \ref{figure: 7}, we can select a highly pure rejuvenating populations by using H$\delta$ EW $<$ 2.3 \AA{} and \DN~$<$ 1.4 in our simple simulation. We adopt these two values as the cutoffs on the \DN-H$\delta$ plane to select potential rejuvenating galaxies in the LEGA-C data. Since we note that active galactic nucleus (AGN) and noisy spectra may contaminate our selections, we remove 5 types of AGN (i.e. MEx \citep{Juneau}, mid-IR, radio, X-ray, and broad-line) and low $S/N$ (i.e. average $S/N < 10$ \AA$^{-1}$) sources from the LEGA-C data, which leaves 1,429 sources with available \DN~and H$\delta$ measurements. In Figure \ref{figure: 8}, we show the distribution of these 1,429 LEGA-C galaxies on the \DN-H$\delta$ plane. The cutoffs that we use are represented by the black lines. In addition, we overplot the contours of Figure \ref{figure: 7} in Figure \ref{figure: 8} to make the comparison of the parameter space better.

As seen in Figure \ref{figure: 8}, there are 14 galaxies within our cutoffs for rejuvenating galaxies, indicating that at least some galaxies matching with these selection criteria exist, though full SED-fitting is required to confirm that these are indeed rejuvenating galaxies. Since the LEGA-C team uses lick indexes \citep{Worthey_94,Worthey_97} rather than the profile fitting to measure the EW, we caution that EW measurement offset between lick indexes and profile fitting will likely need to be addressed in order to ensure robust selection. Furthermore, we also note that superpositions of unrelated objects on the sky can produce potential contaminates. For example, the superposition of a star-forming galaxy and a quiescent galaxies can mimic the spectral feature of a rejuvenating galaxy. In spite of the above potential contaminates, we do not expect many rejuvenating galaxies at z=0.7 as our discussion in Section \ref{sec:intro}. The confirmation of even one rejuvenating galaxy or no rejuvenating galaxies in the LEGA-C data would be valuable for understanding the star-formation regulation process in a cosmological context. In a future study, we will take a detailed investigation on the rejuvenating candidates in the LEGA-C data, and report on the purity and completeness of our selection criteria, by fitting the LEGA-C data with a nonparametric SFH (Nersesian et al. in prep).

\section{Discussion} \label{sec:disc}
\subsection{Searching for rejuvenating systems in large-scale spectroscopic surveys}\label{subsec: large survey}
In the previous sections, we demonstrate that rejuvenating galaxies tend to have small \DN{} features and weak Balmer absorption lines. Here we will describe how to efficiently find such objects in large spectroscopic surveys. 

The main challenge for characterizing the frequency of rejuvenating galaxies with our technique is emission line infilling, which occurs because Balmer line emission in star-forming galaxies will typically render the underlying absorption inaccessible. This is an important limitation because separating rejuvenating and star-forming galaxies requires measuring their Balmer absorption depths to within an accuracy of $\sim1$ \AA~as demonstrated in this work, but these absorption lines are often buried under emission lines with width $\sim5-50$ \AA. Measuring and removing this nebular emission is relatively easier to deal with by using the medium-resolution VLT/VIMOS LEGA-C spectra $R \sim$ 3500) because the spectra are sufficiently high resolution as to be able to separate measurements of Balmer absorption and emission, but harder in more typical moderate resolution surveys (e.g., Sloan Digital Sky Survey (SDSS, $R$ = 1500 at 3800 \AA{} and = 2500 at 9000 \AA{}, \citealt{SDSS_2022})), where there is more blending. Crucially, these moderate-resolution surveys cover larger volumes needed to give the best constraints on number densities of rejuvenating galaxies. One promising upcoming survey is the Prime Focus Spectrograph survey (PFS, $R$ = 2300-5000, \citealt{Greene_2022}), which covers a wide area and moderate spectral resolution, though the S/N for most exposures will not be as high as LEGA-C. Another promising option is the Dark Energy Spectroscopic Instrument survey (DESI, $R$ = 2000-5500, \citealt{Amir_2016}), which is ongoing now. 

One possible solution to the challenge is employing the Balmer decrement (\citealt{Calzetti_2000,Brinchmann_2004,Garn_2010,Price_2014}) (e.g., H$\alpha$/H$\beta$) to estimate the emission line fluxes. Since the atomic properties of hydrogen are well understood, the relative strengths of the Balmer emission lines can be well determined if we know the number of ionizing photons. We can tell the amount of reddening due to dust from the ratio of two Balmer emission line luminosities, as their intrinsic luminosity ratio can be calculated from theory. If we assume a dust reddening curve we can then obtain the dust attenuation at any given wavelength. Then we can predict the integrated flux in any of the Balmer emission lines. Finally, the Balmer absorption line fluxes can be calculated by subtracting the emission line fluxes from the spectra. 

Spectral flux calibration problems can make long-wavelength baselines challenging to interpret (e.g., H$\alpha$/H$\beta$). However, this can be mitigated by using adjacent emission lines (e.g., H$\delta$/H$\gamma$), though the shorter wavelength baseline here makes the relationship to dust attenuation more subtle.

Ultimately, this technique would rely on a tradeoff between a flux-calibration-sensitive long-baseline and a less dust-sensitive short baseline. Therefore, our technique can also be applied to surveys that have relatively deep spectra but not high resolution, if they have a wide range of wavelengths to include as many Balmer lines as possible and have continuum detected to calculate EWs.

\subsection{Effect of Stellar Population Synthesis (SPS) Models on EW Measurement}\label{subsec: spsmodel}
The choice of SPS models can have an impact on the value of EW as a function of age/metallicity of the SSP. \cite{Groves_2012} use different stellar population synthesis models to investigate the dependence of Balmer absorption line EWs on age and metallicity. Their results mostly agree with our Figure \ref{figure: 4}, which conclude that the Balmer absorption line EWs show a strong dependence on age but weak dependence on metallicity. It is also consistent with our results that all of their Balmer absorption line EWs peak at 0.5 - 1 Gyr and that the peak age varies slightly with metallicity. 

However, we notice that the ranges of EWs shown in \cite{Groves_2012} are different from ours. Their results show peak EWs are typically 20\% - 70\% larger than our results, with the larger difference occuring at younger ages. We cannot establish whether this is due to the methodology of measuring EWs or the choice of SPS models. That being said, in \cite{Groves_2012}, they find the variation between SPS models is at the 2x level at fixed age/metallicity, which is likely consistent with variation between SPS models. Therefore, when applying our technique, it will be important to empirically calibrate the EW threshold for rejuvenating galaxies by comparing to independent techniques for identifying these objects, as discussed in the next section.

\subsection{Comparison to other techniques for identifying rejuvenating galaxies}
There are several other methods proposed in the literature to identify rejuvenating galaxies. One possible way is to identify rejuvenation events via measuring the spatial distribution of metallicity within early-type galaxies, as early-type galaxies with recent star formation are found to be more metal-poor than quiescent early-type galaxies due to their metal-poor young stellar populations (i.e., \citealt{Werle_2020,Jeong_2022}). However, this approach only identifies a certain type of rejuvenation, i.e., rejuvenation triggered by metal-poor gas inflow.

Another widely used method is nonparametric SFH analysis (e.g., \citealt{Chauke_2019, woodrum_2022}), where the SFH of galaxies is directly inferred via SED-fitting. This is the most direct approach, but is often computationally expensive because SED-fitting includes many parameters accounting for the complex physics in galaxies. Moreover, results of SED-fitting are sensitive to the adopted priors, which are not always tuned to find rejuvenating systems \citep{Leja_2019a}. Furthermore, \citet{Akhshik_2021} mention that rejuvenation events can occur in both the outskirts and the center of galaxies, depending on the triggers of the rejuvenation process. Hence, if rejuvenation occurs in the center, a spatially resolved spectrum is required.

This study shows that rest frame spectral coverage of 4000-5000 \AA{} proves highly efficient at identifying rejuvenating systems with a nonparametric SFH analysis. Nonparametric SFH analysis can help confirm the ``quick" results from simply measuring Balmer EWs; our method can give us a useful heuristic and fast way to identify these systems. As a result, our technique complements SFH inference by showing the Balmer lines are crucial for identifying systems. 

Nevertheless, our method has some limitations. For example, as we discuss in Section \ref{subsec: large survey}, measuring Balmer absorption line EWs is best performed at sufficient resolution to resolve emission within absorption lines. Low resolution can make it hard to separate absorption lines from emission lines and so reduces the accuracy of Balmer absorption line EWs measurements. This limitation can be mitigated if the survey includes multiple Balmer emission lines and the continuum is detected. 

As we show in Figure \ref{figure: 5}, since strong rejuvenating galaxies resemble star-forming galaxies experiencing mild rejuvenation in the Balmer absorption lines EWs, our methods are most efficient in the selection of mild rejuvenating galaxies. Therefore, the completeness of selection could be a concern if we require a high purity of strong rejuvenating galaxies. However, we note that \citet{Alarcon_2022} show that both the UniverseMachine and IllustrisTNG predict many mild rejuvenation events whereas strong ones are rare. If mild rejuvenation events also dominate in the real universe, our inability to identify strongly rejuvenating systems would be significantly less important.

\subsection{Contrast in the Distribution of Rejuvenating Galaxies between Simulations and Real Universe}
In our simulated populations, we assume equal numbers of rejuvenating galaxies and star-forming galaxies, and thus purity/completeness calculations performed on this population implicitly assume these two galaxy populations have the same number density. However, as we discuss in Section \ref{sec:intro}, the frequency of rejuvenation is very low in the real universe. If there are 20 times more star-forming galaxies then most ``rejuvenating" candidates selected from our technique will be false ones with a EW criterion of high completeness. Therefore, the real distribution of rejuvenating galaxies in the universe can be a concern for our technique. To examine how the real distribution of rejuvenating galaxies in the universe affects our technique, we reduce the percentage of rejuvenating galaxies in our simulation. We find that when the percentage of the rejuvenating population is reduced to $\sim$ 4\% of the total number of star-forming galaxies, the density of the rejuvenating population becomes comparable with the density in the wings of the star-forming distribution. This means that there may be more star-forming contaminants in the real Universe; however, this is only a rough prediction due to the unknown true distribution of recent star formation histories in star-forming galaxies. A future step is to calibrate our purity and completeness limits by estimating observed number densities with nonparametric SFH analyses.

\subsection{The Effect of Extra Dust Attenuation Around Young Star-forming Regions}\label{subsec: Dust1}
In our simulated populations, we employed a two-component dust model \citep{Charlot_2000} to compute the effects of dust on the integrated spectral properties of galaxies. In brief, dust has two components in our model: one is the optical depth of the diffuse dust (i.e., $\hat{\tau}_{2}$) and the other is the optical depth of the birth cloud dust (i.e., $\hat{\tau}_{1}$). Young stars are attenuated not only by dust in the ambient interstellar medium, but also by dust in their birth clouds. It is worth noting that in our model, $\hat{\tau}_{2}$ is a free parameter and $\hat{\tau}_{1}$ is sampled from another free parameter, dust ratio (i.e., $\frac{\hat{\tau}_{1}}{\hat{\tau}_{2}}$). Therefore, $\hat{\tau}_{1}$ is not fixed in our model.

\begin{figure*}[htb]
    \centering
    \includegraphics[width=0.9\textwidth]{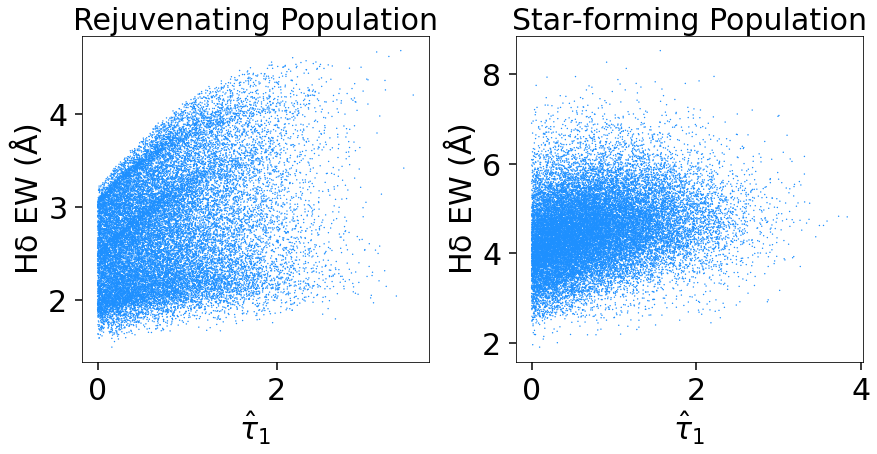}
    \caption{H$\delta$ EW for simulated rejuvenating and star-forming populations as a function of the amount of birth-cloud dust (i.e. $\hat{\tau}_{1}$). The H$\delta$ EWs of the rejuvenating population increase with $\hat{\tau}_{1}$, while the trend for the star-forming population is much weaker, for reasons discussed in \ref{subsec: Dust1}. Increasing levels of extra dust around young stars in galaxies has the potential to marginally decrease the purity of the Balmer EW selection of rejuvenating systems.}
    \label{figure: 9}
\end{figure*}

We notice that $\hat{\tau}_{1}$ has impacts on the EW in our model. In Figure \ref{figure: 9}, we show the variation of EW of H$\delta$ as a function of $\hat{\tau}_{1}$ for both the simulated rejuvenating and star-forming populations. We find a Spearman coefficient $\sim$ 0.274 for the rejuvenating system, which increases to $\sim$ 0.477 if we require a high $\hat{\tau}_{1}$ and EW, while the star-forming system has a Spearman coefficient $\sim$ 0.117. This result indicates that the EW of $H\delta$ is strongly correlated with $\hat{\tau}_{1}$ in the simulated rejuvenating system, where the correlation becomes stronger at high $\hat{\tau}_{1}$ and EW, while the correlation is much weaker in the simulated star-forming system.

This is due to the fact that $\hat{\tau}_{1}$ only affects the youngest stars, which are the main sources of continuum in the simulated rejuvenating population, whereas the older stars which contribute the dominant share of the Balmer absorption are unaffected by this birth-cloud dust. Thus, larger $\hat{\tau}_{1}$ will increase the Balmer absorption line EWs for rejuvenating population. However, for the normal star-forming population, the continuum has significant contributions from older stellar populations as well, and therefore, their EWs are marginally less affected by $\hat{\tau}_{1}$ increment. On the contrary, since the diffuse dust affects both the young and old stars, $\hat{\tau}_{2}$ has no impact on the EW.

This result implies that large $\hat{\tau}_{1}$ will reduce the EW contrast between rejuvenating galaxies and star-forming galaxies, thus decreasing the purity of our technique. However, we caution that the birth cloud dust model is only a rough approximation for the complex star-dust geometry in typical galaxies (see \citealt{Viaene_2016,Bianchi_2018,Narayanan_2018,Nersesian_2019} for a review), and so while there is significant observational support for extra attenuation towards young stars, (e.g., \citealt{Calzetti_1994, Price_2014, Naven_2015}) in detail the boundary between young and old stars in our predictive model may behave quite differently in the real universe. Furthermore, the amount of dust in rejuvenating systems and star-forming system may be different in the real universe than our simple model; this question can be answered once we more closely examine candidates in a large survey.

\section{Conclusion} \label{sec:highlight}
In this work, we investigate the observational characteristics of simulated rejuvenating and star-forming populations at $z = 0.7$ using the modified Prospector$-\alpha$ SED model. We first demonstrate that the star-forming and rejuvenating galaxies have nearly identical broadband colors in our simulation, suggesting that color-color diagrams cannot distinguish between the two types of galaxies. Next, we show that the model rejuvenating galaxies have weaker Balmer absorption lines than the model star-forming galaxies. This is because rejuvenating galaxies had recently been quenched, which leaves them with fewer A-type stars which are distinguished by strong Balmer absorption lines, so called ``O - A" galaxies.

We find that there is a small fraction of star-forming contaminants that have weak Balmer absorption lines similar to those of the rejuvenating population, visible in the bottom panel of Figure \ref{figure: 3}. We look into these contaminants and come to the conclusion that star-forming galaxies which undergo weak rejuvenation, without in fact having experienced quiescence, have similar Balmer absorption lines EW to rejuvenating galaxies that undergo strong rejuvenation. Therefore, we claim that Balmer absorption line EWs are primarily efficient in identifying weak or mild rejuvenating galaxies. 

In addition to the weak Balmer absorption lines, a region in the \DN-H$\delta$ plane is found to be almost completely populated by the rejuvenating galaxies. The rejuvenating population exhibits weak Balmer absorption lines due to recent quenching, and has low \DN~due to relatively high ongoing star formation rate. Figure \ref{figure: 7} demonstrates that, at least in our simulation, rejuvenating galaxies can be chosen with reasonable purity by placing suitable cutoffs in the \DN-H$\delta$ plane.

Based on the above results, this study offers an efficient technique to identify rejuvenating populations through measuring Balmer absorption line EWs. Moreover, this technique provides an intuitive guide for the expected outcome of SED-fitting results. Understanding this technique will also help in survey design such as what range of wavelengths should be detected to better observe rejuvenating populations. However, our technique also has some limitations. First, high resolution spectral data like LEGA-C are required to separate the Balmer absorption from the Balmer emission. If a survey does not have a high enough resolution but measures multiple Balmer emission lines and the continuum, Balmer decrements can also be employed to calculate Balmer emission EWs and thus the absorption EWs. Second, as we discuss herein, the proposed technique is primarily efficient at selecting mild rejuvenating galaxies. 

Although both Balmer absorption lines EW and the \DN-H$\delta$ plane exhibit high purity selection criteria in our models, purity resulting from these criteria are yet to be tested on real data. In order to confirm whether the candidates are actually rejuvenating galaxies, detailed SED-fitting is still essential for measuring metallicities and reconstructing the detailed SFHs.

A logical next step is to use this technique to estimate the number density of rejuvenating systems in LEGA-C survey, and to characterize their detailed SFH and abundances. It will additionally be useful to compare the results from the quick-look observational selection criteria presented here to results from detailed SFH modeling from the LEGA-C survey.

\begin{acknowledgements}
We thank the anonymous referee for their helpful comments, which greatly improved the paper. Computations for this research were performed on the Pennsylvania State University’s Institute for Computational and Data Sciences’ Roar supercomputer.
\end{acknowledgements}

\bibliography{sample631}{}
\bibliographystyle{aasjournal}

\end{document}